\documentclass{article}
\usepackage{spconf,amsmath,graphicx}
\usepackage{booktabs}
\usepackage{amssymb}
\usepackage{float}
\usepackage{comment}
\usepackage{color}

\usepackage{xcolor}
\usepackage{url}

\title{Improving curriculum learning for target speaker extraction with synthetic speakers}
\address{
  $^1$National Institute of Informatics, Tokyo, Japan\\
  $^2$Sokendai, Kanagawa, Japan \\
\texttt{\{yunliu, xuecliu, jyamagis\}@nii.ac.jp}
}
\name{Yun Liu$^{1,2}$, Xuechen Liu$^{1}$, Junichi Yamagishi$^{1,2}$\thanks{This study is partially supported by MEXT KAKENHI Grants (24K21324) and JST, the establishment of university fellowships towards the creation of science technology innovation (JPMJFS2136).}}
\begin{document}
\maketitle
\begin{abstract}
Target speaker extraction (TSE) aims to isolate individual speaker voices from complex speech environments. The effectiveness of TSE systems is often compromised when the speaker characteristics are similar to each other. Recent research has introduced curriculum learning (CL), in which TSE models are trained incrementally on speech samples of increasing complexity. In CL training, the model is first trained on samples with low speaker similarity between the target and interference speakers, and then on samples with high speaker similarity. To further improve CL, this paper uses a $k$-nearest neighbor-based voice conversion method to simulate and generate speech of diverse interference speakers, and then uses the generated data as part of the CL. Experiments demonstrate that training data based on synthetic speakers can effectively enhance the model's capabilities and significantly improve the performance of multiple TSE systems. 
\end{abstract}
\section{Introduction}
\label{sec:intro}

Target speaker extraction (TSE) aims to isolate the voice of a specific speaker from a mixture of voices from various speakers and sounds, including background noise. This technology is crucial in enhancing the quality of tele-commuting, improving the functionality of hearing aids, and refining automatic speech recognition (ASR) systems. Despite advancements facilitated by deep neural networks~\cite{Zmolikova_Spkbeam_STSP19,Wang2019}, TSE continues to face significant challenges, particularly when the speakers have similar vocal characteristics. This similarity makes the extraction process more complex and error-prone, highlighting the need for effective training models.


Simulated audio mixtures are often employed to training neural TSE. Such mixed speech is usually created by directly adding the speech of the isolated target speaker to the various interference speakers. TSE models are then trained to solve the inverse problem of estimating a single speaker's waveform from the mixed speech. Given the diversity of speaker characteristics, it is crucial that a wide variety of speakers are included in the training data as interference speakers. However, the Librispeech~\cite{panayotov2015librispeech} and WSJ0~\cite{wsj0} databases, which are standard databases for TSE tasks, contain a limited number of speakers, which is insufficient to account for the diversity of speaker characteristics. Speaker recognition models typically use data from several thousand speakers. However, the quality of this data is often not clean enough for effective TSE training.
 

One way to generate large amounts of data with such diverse speaker characteristics is to use speech generative models. Significant advances in speech generative models have made it possible to generate speech with human-level naturalness and speaker similarity~\cite{ju2024naturalspeech,casanova2022yourtts,wang2023neural,baas2023knnvc,Lv2023SALTDS}. This situation raises an important question of whether such generative models can be used to generate specialized training data for target speaker extraction. There are various possible strategies for generating training data using generative models, \textit{one of which is to generate diverse synthetic interference speakers from given interference speakers' audio that are different from given interference speakers}.

\begin{figure}[t]
    \centering
    \includegraphics[width=\columnwidth]{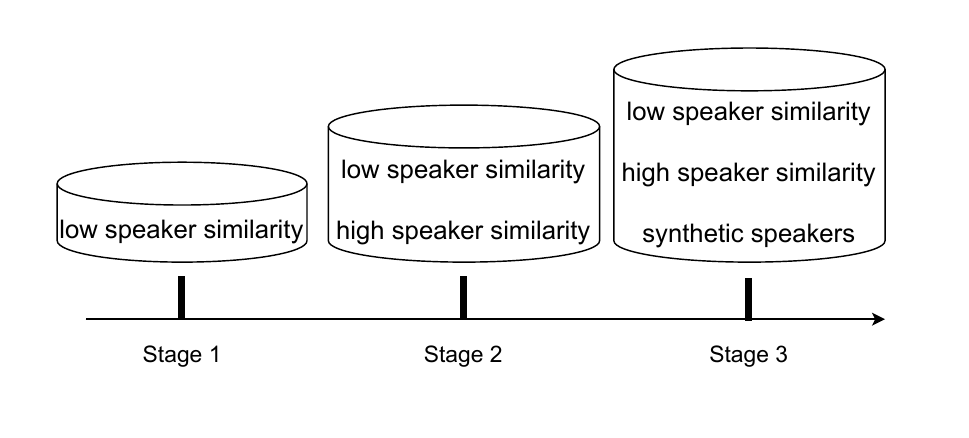} 
    \vspace{-13mm}
    \caption{Three stage curriculum learning}
    \label{fig:stage}
    \vspace{-5mm}
\end{figure}

In this paper, we propose the use of generated synthetic interference speakers in the learning procedure of a recently proposed curriculum learning (CL)-based TSE ~\cite{liu2024target} in order to answer the above scientific question. The CL-based TSE starts with training on 'easy' data samples and progressively schedules the training data to include 'harder' training samples~\cite{wang2021survey}. The previous study ~\cite{liu2024target} reported that speaker similarity, which is measured by cosine similarity between target and interference speakers, serves as an effective difficulty measure when selecting and scheduling training data.  Therefore, we introduce a new curriculum of much more complex training data using synthetic interference speakers based on voice conversion (VC) as well as real  interference speakers. The VC framework used is a recently proposed $k$-nearest neighbor ($k$-NN)-based VC \cite{baas2023knnvc,Lv2023SALTDS}. This strategy is expected to enrich the training set and improve the performance of various TSE systems.



 \begin{figure*}[t]
    \centering
    \includegraphics[width=0.97\textwidth]{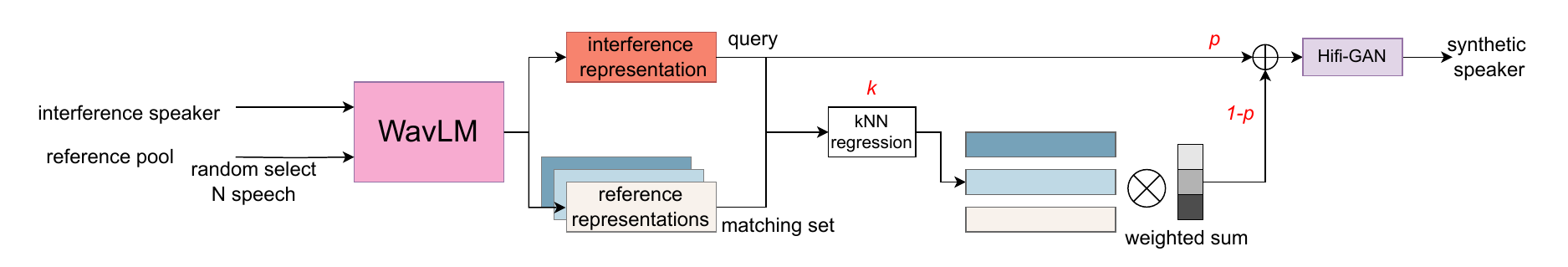} 
    \vspace{-3mm}
    \caption{Synthetic speaker generation using the $k$-NN VC / SALT system.}
    \label{fig:salt}
    \vspace{-3mm}
\end{figure*}

The remainder of this paper is structured as follows: Section 2 reviews  related works. Section 3 describes the proposed method for generating new speakers via the speech generative model. Section 4 presents the experimental settings and results, and Section 5 concludes our findings with future work.


\section{Related work}

\subsection{Designing effective training data for TSE}

Many recent studies have aimed to make the training data for TSE or speech enhancement more realistic and useful. For example, LibriCSS~\cite{chen2020continuous} achieves  more realistic training data by recording overlapped speech in a real setting, using multiple loudspeakers at different positions to play audio from a single speaker. Another study~\cite{Realdata2024} suggests directly using real training data with supervised learning for speech enhancement models to potentially reduce mismatches between training and inference. Moreover, separation models trained with data augmentation~\cite{alex2023data} generalize better to unseen conditions and also enhance the robustness of the model. 
\cite{kuznetsova2023potential} used neural speech synthesis-based data augmentation for personalized speech enhancement to address the typical lack of personal data.
However, these methods randomly select all of the data for training without considering the strategic use of real or synthetic speakers.

\subsection{CL for TSE}

Such synthetic interference speakers can be used from the beginning of training, but if they are too diverse, they may be too complex for a TSE model that has not been well trained.
CL for TSE \cite{liu2024target} is also relevant in this sense. As illustrated in Figure \ref{fig:stage}, in Stage 1 of the CL, the training data consists of target and interference speaker pairs that exhibit low similarity, identified using a speaker encoder. This first stage focuses on easier tasks to establish foundational skills in identifying speaker characteristics. In Stage 2, the model is then exposed to speaker pairs with higher similarity. It has been reported that performance can be improved by using this two-step curriculum.


\section{Improved CL using synthetic speakers}

\subsection{Main concept}

The concept of our method proposed in this paper involves adding a new training stage to the CL framework for TSE, as proposed by Liu et al. \cite{liu2024target}. This new stage, designated as Stage 3 and illustrated in Figure \ref{fig:stage}
, specifically involves using pairs of a real target speaker with a synthetic interference speaker, as well as pairs of real target and real interference speakers.
The increased diversity of interference speakers is expected to further strengthen the models enhanced by the two stages of CL.



\subsection{Synthetic speaker generation based on VC}
\subsubsection{Outline}
Our goal is to use a speech generative model to generate a set of 
diverse synthetic interference speakers that differ from a given set of interference speakers. The speech generative model may either be VC or text-to-speech (TTS). However, we chose not to use a TTS model because we wanted to preserve the original content information in the interference speech.

After initial experiments, we selected $k$-NN VC~\cite{baas2023knnvc} and its application to a speaker anonymization system~\cite{Lv2023SALTDS} called SALT\footnote{https://github.com/BakerBunker/SALT} as our speaker generation model.  As Figure \ref{fig:salt} shows, SALT comprises three primary components: a pre-trained self-supervised latent encoder based on WavLM~\cite{chen2022wavlm}, a $k$-NN regression module, and a vocoder based on HiFi-GAN~\cite{kong2020hifi}. 
It performs speaker anonymization by first randomly selecting $N$ reference speakers in the reference pool.
Then probabilistic weighting is applied to the WavLM representations of $k$ neighbors chosen from these $N$ reference speakers, using the $k$-NN algorithm where the input speaker’s representation serves as a query. The input speaker's representations are replaced with this weighted summed representation. We use this system to generate many synthetic interference speakers.

Note that in the WaveLM-based encoder, only one specific layer's (the third  layer) features that are strongly correlated with the speaker information are extracted. This produces a single 1024-dimensional vector for every 20 ms. The vocoder is also trained using the specific layer's features as input features.

\subsubsection{Generation of synthetic interference speakers}

Here, we describe how we use the SALT system to  generate  synthetic interference speakers. Let $s_{t}$ be a given interference speaker's audio of TSE.  We first extract the input speaker's WavLM representation, $\text{WavLM}(s_t)$, of the audio $s_{t}$. Using the extracted representation $\text{WavLM}(s_t)$ as a query, the basic  $k$-NN is used at each time step to select $k$ neighbors from the representations of the $N$ reference speakers. The cosine distance is employed to determine the nearest neighbors in the $k$-NN process. 


After selecting $k$ neighbors, we assign a random weight $w_j$ to each selected neighbor's WaveLM representation, $\mathbf{R}_j$. These weights are sampled from a normal distribution and then normalized so that their sum equals 1. A weighted sum of these representations is then calculated. This sum is blended with the original interference speaker's representation through linear interpolation to generate a synthetic speaker similar to the input speaker, as follows: 
\begin{align}
(\mathbf{R}_1, \cdots, \mathbf{R}_k) &= \text{kNN} (\text{WavLM}(s_t))\\
\mathbf{O} &= (1 - p) \sum_{j=1}^k w_{j} \mathbf{R}_{j} + p \text{WavLM}(s_t) 
\end{align}
where $p$ is a hyperparameter that needs to be adjusted. We then input $\mathbf{O}$ into the HiFiGAN vocoder to generate a waveform.

This framework is ideal for generating synthetic interference speakers. By adjusting the parameters $k$ and $p$, it is possible to create a large and diverse array of new interference speakers that are distinct from the existing set. Additionally, the number of interference speakers can be expanded by generating as many random weights as needed.

\section{Experiments}
\label{sec:typestyle}

\subsection{Dataset}
We conducted our proof-of-concept experiments using the Libri2talker dataset~\cite{xu2021target}, a 2-talker simulated mixture audio dataset derived from LibriSpeech~\cite{panayotov2015librispeech}. Compared with Libri2Mix~\cite{cosentino2020librimix}, modifications were made to the dataset by alternating the target and interference speakers. Additionally, each 2-talker mixture was utilized twice, consistent with the experimental setup in the previous study~\cite{xu2021target}.


\subsection{Feature extraction}
The sampling frequency of the speech waveform was set to 16 kHz. The reference speech  samples normalized to a length of 15 seconds, either by padding or segmenting, while all mixture speeches were clipped to 6 seconds. The STFT parameters for the mixture signal included a window length of 32 ms, a hop size of 8 ms, and an FFT length of 512.

\subsection{Hyperparameters}
Training involved an Adam optimizer, initiated with a learning rate of \(1 \times 10^{-3}\). Following an adaptive scheme, the learning rate was increased over the first 5,000 batches and then gradually reduced to ensure it did not fall below \(1 \times 10^{-5}\). Each model underwent three training iterations with different random seeds, and the results were average.  We used the early stopping method, stopping training if the dev set results did not improve for six consecutive epochs. The model with the highest iSDR on the dev set was selected for the final evaluation.

\subsection{Model setting}
For TSE, we utilize the Conformer 4-layer model described in ~\cite{liu2024target} as our main TSE model. In addition, BLSTM, SpeakerBeam~\cite{SpeakerBeam2019} and VoiceFilter-based TSE models~\cite{wang2018voicefilter} were also trained. 

The main Conformer-based TSE model estimates complex masking ratios~\cite{cirm} for the real and imaginary parts of a complex-valued spectrum. The input sequence consists of a time-domain waveform of the mixed signal and features from the mixture are then concatenated with the speaker embeddings extracted by the speaker encoder. These embeddings were obtained using a pre-trained ECAPA-TDNN model~\cite{desplanques2020ecapa} that was pre-trained on the VoxCeleb2 dataset~\cite{chung2018voxceleb2}. 

For VC, we utilize the WavLM-Base model trained on the LibriSpeech dataset and extract the latent space representations from the third layer. As the reference pool, we randomly selected $N=20$ speakers from the LibriSpeech train-clean-100 dataset, which was also used as the training set for the vocoder. This means that the VC model also used the same training set, LibriSpeech, as the TSE model, without introducing any new training data. For each selected speaker, we randomly choose 50 audio samples to extract features. 

\subsection{Synthetic data generation setting}
The Libri2talker dataset contains 127k data triplets for training, each including a mixture, target speech, and reference speech. 

For each data triplet in the training set, we input the interference speech (mixture-target speech) into the SALT system to create a new different interference speaker. 
For each combination of $k$ and $p$, we simply generate 127k synthetic speakers offline unless stated otherwise.  The $k$ value is set to 4, 8, or 20. The interpolation factor $p$ is set to 0.2, 0.5, or 0.8. 

To incorporate both real and synthetic interference speakers in the training process, we set the batch size of each mini-batch to 48. Each mini-batch includes an equal number of real and synthetic speakers.

\subsection{Evaluation metric}
For the evaluation metric, we use improved signal-to-distortion ratio (iSDR) in dB, which is commonly used to validate TSE, speech enhancement, and speech separation. The iSDR is measured as the relative increase in SDR compared to the original mixture.  A higher score indicates higher model performance.

\label{sec:pagestyle}
\subsection{Results}

\begin{table}[t]
\centering
\caption{Experimental results before and after adding Stage 3, which uses synthetic speakers, to the TSE model based on Conformer in \cite{liu2024target}.
 Hyperparameters are $k$=4 and $p$=0.5.}
\label{tab:allmethods}
\begin{tabular}{cccc}
\toprule
\textbf{TSE method} & \textbf{Stage2} & \textbf{Stage3} & \textbf{iSDR [dB]} \\ \hline
Conformer w/o CL~\cite{liu2024target} & $\times$ & $\times$ &  12.57 \\ 
Conformer~\cite{liu2024target} & $\checkmark$ & $\times$  & 13.44 \\ 
Conformer (real only) & $\checkmark$ & $\checkmark$ &  13.98  \\ 
Conformer (real+syn) & $\checkmark$ & $\checkmark$ & \textbf{14.43} \\ 
\bottomrule
\end{tabular}
\end{table}

Table \ref{tab:allmethods} shows the experimental results of using the main TSE model based on the Conformer. First, the Conformer (real only) configuration, which involves simply adding Stage 3 to the curriculum proposed by Liu et al.~\cite{liu2024target} and continuing training without synthetic speakers, shows an improvement in iSDR from 13.44 dB to 13.98 dB. Next, the Conformer (real+syn) configuration, which incorporates synthetic interference speakers in Stage 3, achieved an iSDR of 14.43 dB. This result not only surpasses the Stage 2 outcome but also exceeds the performance of training without synthetic speakers. These findings clearly demonstrate the benefits of integrating synthetic interference speakers into TSE training.

\subsection{Analysis}

This subsection analyzes some of the questions that arise in using the proposed method.

\noindent 
\textbf{Q1: Should real and generated speakers be used together?}

\begin{figure}[t]
    \centering
    \includegraphics[width=0.85\columnwidth]{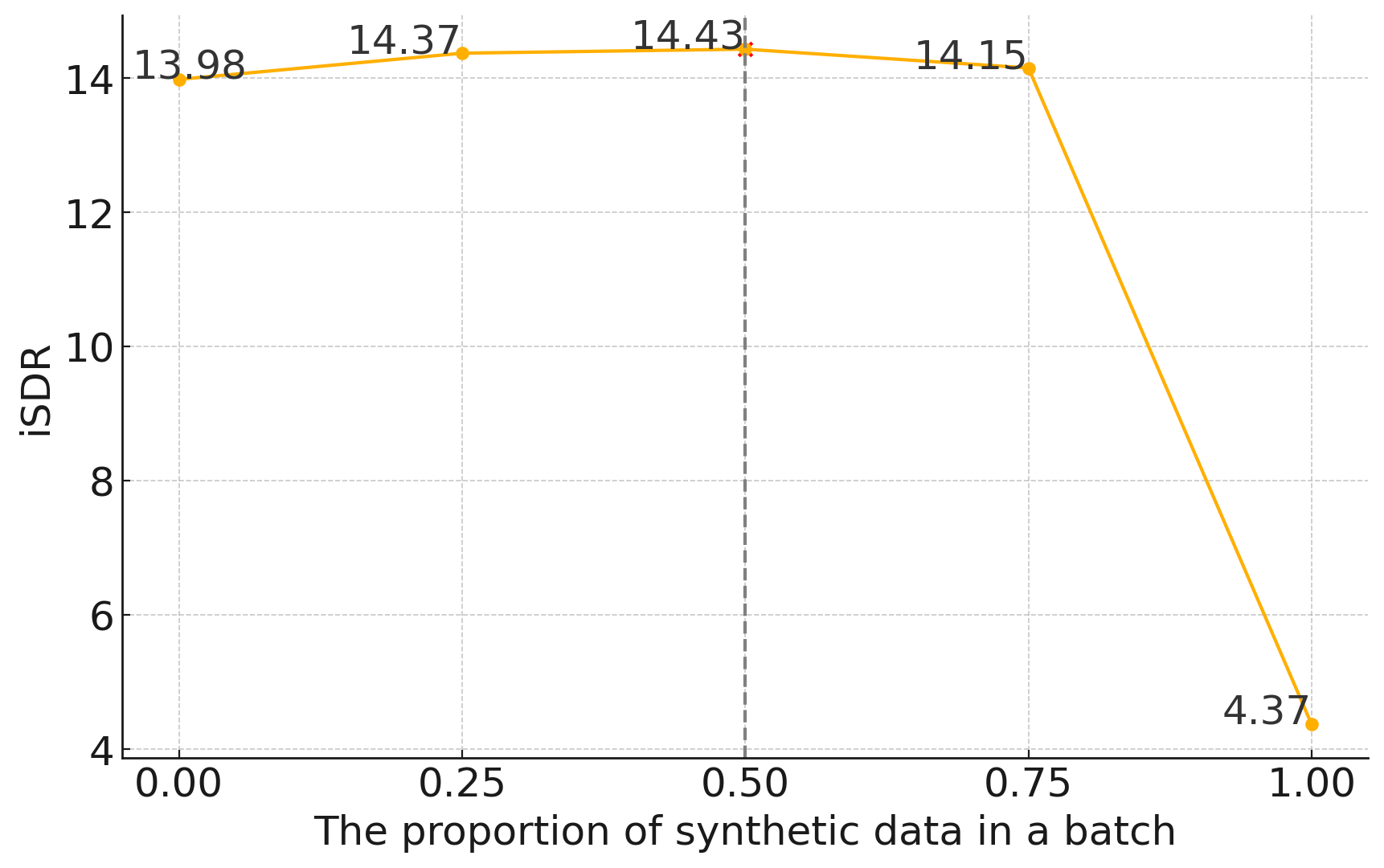} 
    \caption{Comparison on the ratio of synthetic speakers within a mini-batch.}
    \label{fig:port}
\end{figure}

To determine whether the synthetic interference speakers alone are sufficient or whether real interference speakers should be used in conjunction, we adjusted the proportion of synthetic interference speakers within each mini batch and compared the performance using the Conformer model. Figure \ref{fig:port} shows the iSDR (dB) performance as a function of the proportion of synthetic interference speakers within a mini batch. An x-axis value of 1  indicates the exclusive use of synthetic interference speakers without real interference speakers. 

\label{sec:task}

The highest iSDR of 14.43 dB was observed at the ratio of 0.50. However, the performance sharply declined to 4.37 dB when the ratio of synthetic speakers increased to 1.00. This significant drop can be attributed to data mismatch as the model, initially trained on real data, is only fine-tuned  with synthetic data, which differs significantly from the real data distribution of the test set.


\noindent 
\textbf{Q2: What is the impact of hyper-parameters used for speaker generation?}

SALT contains  two hyperparameters:  $k$, which represents represents the number of selected $k$-nearest neighbors, and $p$, which dictates  the preservation factor of retaining the original interference speaker. Here, we investigate the impact of these two hyperparameters. Table \ref{tab:kp} displays the iSDR (dB) outcomes across various configurations of the parameters $k$ and $p$.

\begin{table}[t]
\centering
\caption{iSDR (dB) results of methods with varying  parameters  $k$ and $p$.}
\begin{tabular}{cccc}
\toprule
\label{tab:kp}
 & $p$=0.2 & $p$=0.5 & $p$=0.8 \\
\hline
$k$=4 & 14.16 & \textbf{14.43} & 14.15 \\
$k$=8 & 14.17 & 14.32 & 14.25 \\
$k$=20 & 14.39 & 14.42 & 14.36 \\
\bottomrule
\end{tabular}
\end{table}

A general trend of improvement in iSDR values is observed as the hyperparameter $k$  increases, demonstrating the benefits of increasing $k$. Particularly noteworthy is the stability of the results at $p=0.5$, where the iSDR values show less variation and maintain higher performance as $k$ changes from 4 to 20. This suggests that a $p$ value of 0.5 is a suitable interpolation factor to use for TSE.

\noindent 
\textbf{Q3: What happens if more interference speakers are generated?}

Previous experiments used the same number of synthetic interference speakers as real speakers during the TSE training. However, it is possible to generate a more diverse set of synthetic speakers, and we investigate this potential here using the following two methods. The first method uses different values of $k$ and $p$, generating an equal number of speakers for each combination and then mixing them. The second method keeps the values of $k$ and $p$, but re-samples the weights $w_j$ to generate more speakers.

\begin{table}[t]
\centering
\caption{iSDR results with different numbers of synthetic speakers.}
\begin{tabular}{lcc}
\toprule
\textbf{TSE dataset}       & \textbf{Data amount} & \textbf{iSDR (dB)} \\
\hline
1:1           & 127k utt.                & 14.43              \\
Different $k$ \& $p$   & 127k $\times$ 5 utt.     & 14.35              \\
Same $k$ \& $p$        & 127k $\times$ 5 utt.     & 14.49              \\
\bottomrule
\label{tab:dataamount}
\end{tabular}
\end{table}
Table \ref{tab:dataamount} shows the iSDR results for experiments that utilized additional synthetic interference speakers. The original Libri2Talker dataset contained 127k utterances. For the ``Different \(k\) \& \(p\)'' condition, we either used \(k=20\) \textbf{or} \(p=0.5\), which expanded the data volume to five times the original amount, comprising five distinct datasets. The ``Same \(k\) \& \(p\)'' configuration consistently used \(k=4\) \textbf{and} \(p=0.5\), but generated five times the amount of data; each dataset was unique due to random variations in the weights assigned to speakers during generation.

The results indicate that merely increasing the volume of synthetic data does not significantly enhance results; both modified conditions produced similar iSDR values around 14.35 dB and 14.49 dB, respectively, compared to 14.43 dB when the ratio was 1:1. This suggests that the 127k utterances seem to suffice for the network to learn the current limited speaker variations of Libri2Talker's test set.

\subsection{Experiments using other architectures}

\begin{table}[t]
\centering
\caption{iSDR results using different TSE architecture. Hyper-parameters for speaker generation are $k$=4 and $p$=0.5.}
\label{tab:netwroks}
\begin{tabular}{crrrr}
\toprule
\textbf{TSE networks} & \textbf{base} & \textbf{real only} & \textbf{real+syn} \\
\hline
Naive-BLSTM & 8.30  & \phantom{1}9.79  & \textbf{10.32}  \\
SpeakerBeam \cite{SpeakerBeam2019} & 9.78   & 10.10  & \textbf{10.51}  \\
VoiceFilter \cite{wang2018voicefilter} & 7.17 & \phantom{1}8.66  & \textbf{9.32}  \\
\bottomrule
\end{tabular}
\end{table}

Table \ref{tab:netwroks} shows the iSDR (dB) for several network architectures, each adapted for specific functionalities. The Naive-BLSTM aligns with the input-output configuration of the Conformer TSE, substituting its four Conformer blocks with a two-layer, 512-unit BLSTM. SpeakerBeam utilizes the established structure from prior research as referenced in \cite{SpeakerBeam2019}. Meanwhile, VoiceFilter modifies the approach from \cite{wang2018voicefilter} by replacing the d-vector with a speaker embedding using ECAPA-TDNN.

In training scenarios, the \textbf{base} category employs curriculum learning with an initial focus on speaker similarity scores under 0.6 for 100 epochs, followed by a  5-epoch finetuning. The \textbf{real} setup fine-tunes exclusively with real data during Stage 3, while the \textbf{real+syn} setup fine-tunes with a mix of real and synthetic data. The proportion of synthetic speaker in each mini batch is also 0.5.

The improvement in iSDR values when using synthetic and real speakers simultaneously highlights the advantages of the proposed approach and demonstrates that it is possible to increase accuracy even with models from different architectures.

\section{Conclusion and future work}
\label{sec:con}

In this paper, we have focused on addressing the limited diversity of speakers in the standard training database for TSE. We proposed a method to generate different synthetic interference speakers from a given set of interference speakers using the $k$-NN VC system and use them as part of the data to create mixtures. The performance of the four TSE models was improved by using both the \textbf{synthetic interference/real target} speaker pairs and the \textbf{real interference/real target speaker} pairs simultaneously in a new stage added to the curriculum learning. It is important to emphasize that this VC system uses the same training dataset as the TSE, meaning no new real data has been added, and the structure of the TSE model has not been altered in any way.

Experiments aimed at increasing the number of interference speakers did not yield significant improvements, suggesting that the diversity of interference speakers in the test set of the standard TSE database may be inherently limited and insufficient. Consequently, a new TSE database that incorporates a broader range of target and interference speakers may be necessary in the future.

Our future work will focus not only on increasing the diversity of interference speakers but on generating interference speakers that are more similar to the target speaker to be separated, as well as producing multi-speaker interference speech.



\bibliographystyle{IEEEbib}
\bibliography{refs}

\end{document}